\documentclass[sn-mathphys,Numbered]{sn-jnl}
\usepackage{geometry}
\usepackage{graphicx}%
\usepackage{multirow}%
\usepackage{amsmath,amssymb,amsfonts}%
\usepackage{amsthm}%
\usepackage{mathrsfs}%
\usepackage[title]{appendix}%
\usepackage{xcolor}%
\usepackage{textcomp}%
\usepackage{manyfoot}%
\usepackage{booktabs}%
\usepackage{algorithm}%
\usepackage{algorithmicx}%
\usepackage{algpseudocode}%
\usepackage{listings}%

\theoremstyle{thmstyleone}%
\theoremstyle{thmstyletwo}%

\theoremstyle{thmstylethree}%

\begin{document}

\title[Rapid Photo-Bleaching of Gamma-Irradiated Yb-doped Optical Fibers by High-Energy Nanosecond Pulsed Laser]{Rapid Photo-Bleaching of Gamma-Irradiated Yb-doped Optical Fibers by High-Energy Nanosecond Pulsed Laser}


\author*[1]{\fnm{Esra} \sur{ Kendir Tekg\"{u}l}}\email{fiz.esrakendir@gmail.com}

\author[1]{\fnm{B\"{u}lend } \sur{Orta\c{c}}}\email{ortac@unam.bilkent.edu.tr }

\affil*[1]{\orgdiv{Bilkent University UNAM-National Nanotechnology Research Center}, \orgname{Institute of Materials Science and Nanotechnology}, \orgaddress{ \postcode{06800}, \state{Ankara}, \country{Turkey}}}




\abstract{ A rapid and efficient photo-bleaching process was demonstrated with a high-energy nanosecond pulse to recover existing and/or revealed color centers on 10 kGy Gamma-irradiated Yb-doped optical fiber. Multi-mJ pulsed laser based on an optical parametric amplifier system operating at wavelengths of 532 nm, 680 nm and 793 nm was used.  The photo-bleaching performance is investigated as a function of the wavelength and energy of the pulsed light source. It was observed that the photo-bleaching level of the Yb-doped optical fiber increased when the exposure time of the pulsed laser light and the photon energy was increased. The results show that the recovery levels of color centers in the Yb-doped optical fibers reached up to 96 \% in a shorter time by using the pulsed laser compared to that of the studies by using the continuous laser.}




\maketitle

\section{Introduction}\label{sec1}
The transmission properties of rare-earth-doped optical fibers can be disrupted in Radiation Induced Attenuation (RIA) and Photo-darkening (PD) processes, which cause a performance loss in the optical fibers. Here, the RIA is an attenuation of the output signals due to radiation-induced defects in optical fibers used in an environment including radiation \cite{Friebele1984,Girard2013}. The PD is a loss mechanism that occurs during high-intensity excitation and laser light generation in high-power Yb-doped fiber lasers and amplifiers due to radiation exposure. This mechanism directly affects the performance of the optical fiber. The PD limits the excitation intensity in fiber lasers and amplifiers and affects the reliability of the system in the long-term \cite{Koponen2007,Koponen2006,Morasse2007}. 
The RIA and the PD are directly related to the formation of color centers in the optical fibers. In the optical fiber structure, the color centers are formed when electrons and holes are produced by an excitation other than radiation and are trapped in a defect in the silica matrix. In the transmitted light into the optical fiber, an absorption at certain wavelengths causes a loss in the signal. The occurrence of RIA and PD occurs due to the number of doped elements (Al, P, etc.), the ratio of Yb$^{+3}$ ions \cite{Koponen2006}, the matrix of elements (Al, P and SiO$_2$) with Yb$^{+3}$, the pumping power, the signal wavelength \cite{Koponen2007} and temperature \cite{Soderlund2009}. Especially, the Yb$^{+3}$ ions and other doped elements may cause an increase in transmission losses. It is a problem that needs to be solved, especially for high-power systems with fiber lasers. Various methods have been proposed in the literature to prevent RIA and PD. 
One of them is the photo-bleaching (PB) \cite{Piccoli2014} method, which is generally formed by the interaction between an optical fiber and a CW light with wavelengths of 355 \cite{Manek-Honninger2007}, 405 \cite{Piccoli2014b}, 532 \cite{Cao2019}, 543 \cite{Guzman2007}, 633 \cite{Gebavi2012} and 793 nm \cite{Zhao2015}. Other methods are H$_2$, or O$_2$ loading method \cite{Yoo2007}, thermal annealing \cite{kuhnhenn2004}, and the adjusting of the ratio of the doped ions such as Al, Ce, P, etc. \cite{Engholm2009}. The PB is a more useful method than the others because the structural change in the optical fiber is not needed. By propagating the light of different wavelengths and different powers into the optical fiber \cite{Firstov2015}, transmission properties of the observed color center absorption bands (at certain wavelengths) are changed and hence the defects are recovered. This recovery mechanism depends on the wavelength and power of the light source, and the type of light source, which consists of photons, excites the oxygen deficiency and/or excess and the bond structures of unpaired electrons and hole defects in the matrix of optical fiber. This means that photons transfer energy to defective areas. Thus, it is achieved by reinstating defective bond structures, or missing ions \cite{Henschel2000,Ghahrizjani2020}. 

Preventing or recovering the formation of such defects before and after the production of optical fiber is of great importance in the production of both efficient and long-lasting systems. To increase the hardness of the Yb-doped optical fibers, the behavior of the elements, which are in the matrix of the Yb-doped fibers, are determined when they expose to gamma radiation or PD. In the literature, few studies are available about Yb-doped optical fibers, and it is known that the maturity of the studies fully reveals the problems and specifies the solutions. Low-power laser sources that produce continuous laser light at a certain and few wavelengths have been used for PB \cite{Borman2016}. In that study, continuous laser light was used on Yb-doped optical fiber for 18.3 hours; after applying a total dose of 1.45 kGy radiation, the recovery of 34 \% and 8 \% were achieved in the transmission of light intensity at 650 nm and 1064 nm, respectively. 
\cite{Borman2016}. In a similar study, Yb-doped optical fiber recovery of 11 \% was observed after 118 hours using continuous laser \cite{Poulin2016}. These studies show that the number of photons per unit of time directly affects the duration and level of the recovery process \cite{Zhao2015,Cao2019}. 

Consequently, the level of recovery increases with increasing photon numbers, and photon enhancement are one of the essential methods in the literature \cite{Zhao2015,Cao2019}.

In the present study, a PB study was performed to improve the existing and/or revealed color centers in a Yb-doped optical fiber after 10 kGy gamma radiation. A pulsed laser with an optical parametric amplifier (OPA) system was used for the PB process. After the pulsed laser process, the results show that the recovery levels of color centers in the Yb doped optical fiber have been achieved to 30 \%-96 \%. In the optical fiber, the maximum PB was observed in the pulsed laser with a wavelength of 532 nm. At the same time, the change of PB was presented with increasing applied energy of the 532nm and 680 nm pulsed laser.

\section{Experimental Procedure}

The PB processes were studied on the well-known commercial Nufern fiber (LMA-YDF-20/400-M). The Nufern fiber has a 20 $\mu$m core and 400 $\mu$m cladding diameter. The core of the optical fiber contains 0.18 mol\%, 1.31 mol\% and 1.42 mol\% of Yb, Al and P elements, respectively. The length of the optical fibers is 2 m \cite{Kendir2022, Akchurin_2019}.

\begin{figure}[!h]
	\centering
	\includegraphics[width=.6\linewidth]{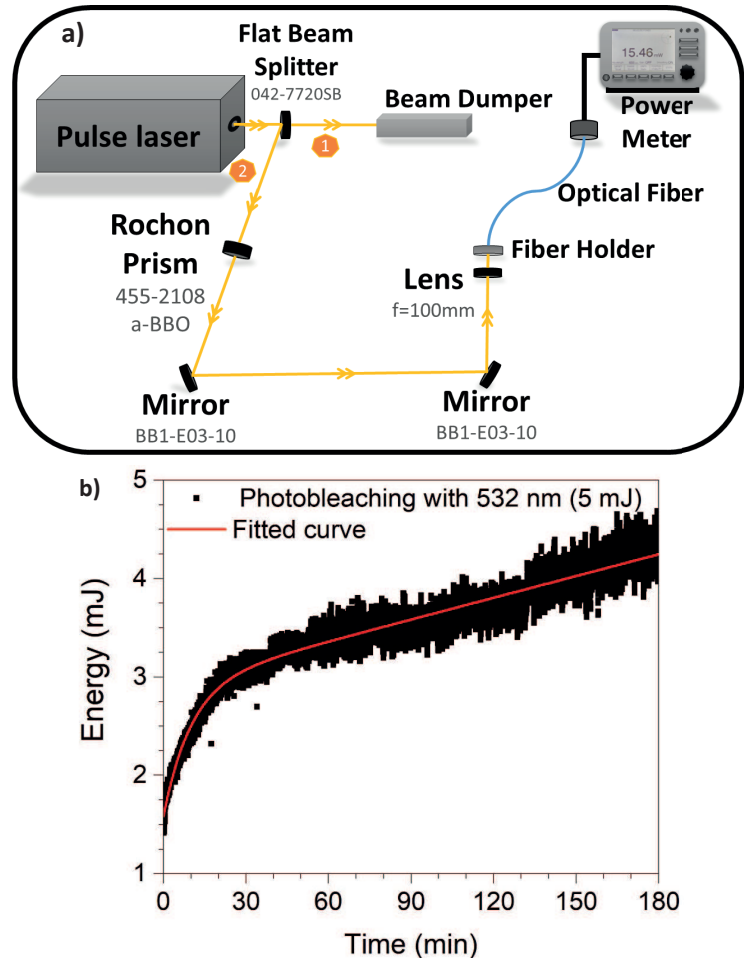}
	\caption{Experimental setup a) the system in which a pulsed laser to Yb-doped optical fibers at wavelengths of 532 nm, 680 nm, and 793 nm. b) The output energy variation of the optical fiber due to the time (Online measurement). The optical fiber, which was irradiated to 10 kGy total doses, was exposed to a 532 nm pulsed laser (5 mJ input energy) for the PB process.
		}\label{fig1}
\end{figure}

Two different experimental systems (offline and online) have been set up for the PB measurement, and these are shown in previous study in Ref.\cite{Kendir2022} and Figure \ref{fig1}.a, respectively. The changes in light intensity of optical fibers were performed before irradiation, after 10 kGy irradiation, and after exposure to pulsed laser. The spectra were recorded as a function of the wavelength using the Xe lamp (Newport Oriel LCS-100) light source and a spectrometer (Ocean Optics-Flame-S-XR1) by means of collimating optics. A measurement procedure was determined to obtain the spectra before and after the irradiation. In the procedure, non-irradiated optical fiber (e.g. non-irradiated Fiber1) was measured, and then the irradiated fiber was performed (e.g. irradiated Fiber1) to compare in the same condition. For each fiber, the procedure was carried out. The system in Figure \ref{fig1}. a was set up to expose 10 kGy irradiated fibers to the light of different wavelengths and energies with a pulse laser. The pulsed laser works at the wavelength range of 660-2400 nm, and its energy range is 120-10 mJ with a frequency of 10 Hz (5 ns). The laser light was reduced to 3 mJ and 5 mJ levels. Because the optical fiber can only carry these energy levels. The energy value entering the optical fiber was reduced by establishing an optical setup with a semi-permeable mirror and polarized cube. The laser light was collimated with an optical lattice system consisting of lenses. The collimated light was coupled to the inside of the optical fiber. The light intensity was measured from the other end of the optical fiber. 

The PB measurements were performed using a pulsed laser system and a Xenon light source in the irradiated optical fibers, which were exposed to 10 kGy gamma radiation. Firstly, the spectra of non-irradiated and irradiated optical fiber were taken in the 400-1000 nm range using the Xenon lamp light. Then, the irradiated optical fibers were exposed to a pulsed laser for 1 hour and their spectra were also taken. Recovery spectra were obtained in optical fibers. These measurements were repeated 3 times. Before each recovery process, the spectrum of the optical fiber was checked. The PB measurements in optical fibers were taken at three different wavelengths of 532 nm, 680 nm, and 793 nm and at 3 mJ and 5 mJ energy values. 

The optical fibers were exposed separately at each wavelength with the pulsed laser for a total of 3 hours. For each wavelength, the light intensity versus wavelength spectrum was recorded. To analyze, four different wavelengths (420 nm, 490 nm, 575 nm, and 650 nm) were selected from the obtained spectra. The RIA values were calculated for these wavelengths Ref\cite{Kendir2022}.

\section{Result and Discussion}

The obtained results indicate that the optical fibers lost 94\% of their transmission properties when exposed to radiation. The RIA spectra show that the defect centers occur in the optical fiber and these defects are specific color centers, which absorb the light, the light intensity inside the fiber losses a large part of it. 
The color centers have been investigated in the literature, and, their wavelengths and full-width-half-maximum (FWHM) values have been indicated in most of the studies \cite{Girard2019,Giacomazzi2018}. Therefore, the expected and known color centers were fitted in the RIA spectra. Fiber's RIA spectra consist of intrinsic color center NBOHC, and extrinsic color centers AlOHC and POHC. The OA bands of the POHC, AlOHC and NBOHC color centers in the optical fiber were found at wavelengths of 400 nm, 466 nm, 536 nm and 664 nm, respectively \cite{Kendir2022}.

Figure \ref{fig1}.b shows the online measurement of the output energy variation from the optical fiber depending on the irradiation time with the pulsed laser. The irradiated Yb-doped optical fiber was exposed to a 532 nm pulsed laser for the PB process. The output energy begins to increase to a 3 mJ value, rapidly, and then the increase slowly continues through a 5 mJ value. The variation in the energy indicates that the Yb-doped optical fiber was recovered due to time. This online measurement proves that rapid and efficient PB performance could be obtained.

\begin{figure*}[!h]
	\centering
	\includegraphics[width=\linewidth]{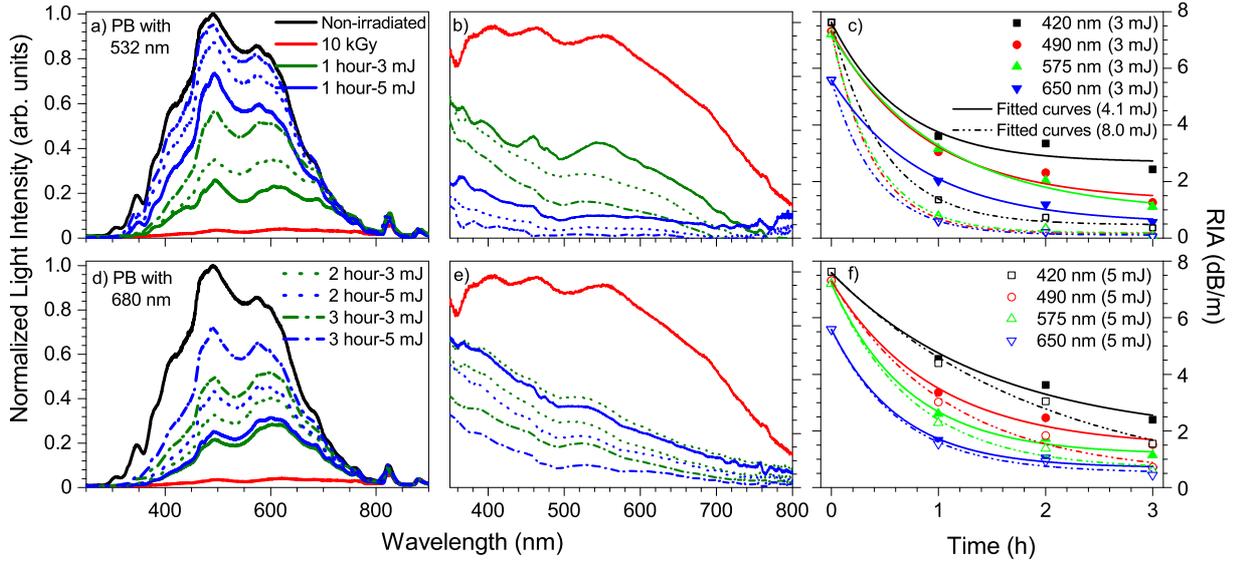}
	\caption{Normalized light intensity change of non-irradiated and 10 kGy irradiated Yb-doped optical fiber after exposure to pulsed laser at 532 nm (a) and 680 nm (d) for 1 hour, 2 hours, and 3 hours. The RIA changes of the fibers (b) and (e). Time-dependent RIA changes at PB wavelengths selected (420 nm, 490 nm, 575 nm ve 650 nm) after healing with a pulsed laser of 532 nm (c) and 680 nm (f).
	}\label{fig2}
\end{figure*}

\begin{table*}[!h]

	\centering
	\caption{The percent recovery values of 10 kGy irradiated Yb-doped optical fiber were calculated after the PB process (after 1 hour, 2 hours, and 3 hours) with 532 nm, 680 nm (energy values of 3 mJ and 5 mJ) and 793 nm (3 mJ) pulsed laser. These percentage values were for selected PB wavelengths of 420 nm, 490 nm, 575 nm, and 650 nm.(Variation of recovery values are percentage)}
	\resizebox{\textwidth}{!}{
	\begin{tabular}{ccccccccccccc}
		\hline
		
\textbf{Selected Wavelengths }                &&\textbf{420 nm}&&&\textbf{490 nm}&&&	\textbf{575 nm}&&&	\textbf{650 nm}& \\
\textbf{Irradiation wavelength}&\textbf{1 hour}&\textbf{2 hour}&\textbf{3 hour}&\textbf{1 hour}&\textbf{2 hour}&\textbf{3 hour}&\textbf{1 hour}&\textbf{2 hour}&\textbf{3 hour}&\textbf{1 hour}&	\textbf{2 hour}&\textbf{3hour} \\
\textbf{532 nm (3 mJ)}	    &  19.6&	22.1&	33.6&	25.4&	35.6&	57.8&	24.2&	40.6&	62.2&	42.7&	62.7&	77.0 \\

\textbf{532 nm (5 mJ)}	    &  53.5&	71.0&	84.4&	72.9&	87.6&	95.5&	69.5&	85.0&	95.5&	75.7&	90.2&	\textbf{96.9} \\

\textbf{680 nm (3 mJ)}&12.3&	18.8&	33.2&	21.3&	32.1&	49.7&	30.0&	45.2&	59.1&	46.2 &	61.5&	74.4 \\
\textbf{680 nm (5 mJ)}&	13.2&	24.5&	48.9&	25.2&	43.4&	72.1&	34.8&	52.8&	76.0&	49.1&	65.0&	81.5 \\

\textbf{793 nm (3 mJ)}    &6.7&	11.8&	12.4&	9.1&	16.8&	19.4&	11.9&	22.0&	28.4&	23.4&	40.5&	\textbf{51.0}  \\

		\hline
	\end{tabular}
}
	\label{tab1}
\end{table*}

 A pulsed laser at three different wavelengths was coupled inside the optical fiber irradiated with the radiation.  The energy of the pulsed laser of 532 nm and 680 nm was 3 mJ and 5 mJ. The energy of the pulsed laser of 793 nm was only 3 mJ. For each wavelength, different samples were used for the PB process. Figure \ref{fig2}.a and d shows the spectra obtained from non-irradiated (black line) and irradiated (red line) optical fibers for PB with 532 nm and 680 nm wavelengths, respectively. The green and blue lines are the PB for 1, 2, and 3 hours. The RIA changes are given in Figure \ref{fig2}.b and e. After the recovery process, the RIA changes decrease with increasing exposure time and energy of pulse laser at selected wavelengths in Figure \ref{fig2}.c and f. The results indicate that the pulsed laser with low wavelength (for 532 nm) carries high energy. Therefore, the defects, which are created by the radiation, can be easily recovered at a certain value by the PB process (The recovered levels are given in Table \ref{tab1}). An electron-related color center is created in Yb$^{2+}$, as well as a hole-related color center such as AlOHC, and NBOHC. The PB is the reverse action of these processes and could eliminate color centers. Hence, Yb$^{2+}$ ion recombines with the released hole of surrounded ligands with the PB (converting Yb$^{2+}$ ion into Yb$^{3+}$ ion) and simultaneously eliminating the hole-related color centers such as AlOHC, and NBOHC \cite{Cao2019,Guzman2007}. It is mean that the energy of the pulse laser light rearranges the bonds between the Yb ions and their neighboring ligands during the healing process of the optical fiber. The decrease in the RIA values of the optical fiber may be related to the rearrangement of the local charge defined as Yb$^{+2}$ $\rightarrow$ Yb$^{+3}$ \cite{Gebavi2012}.

For long wavelengths (for 793 nm), it was observed that the light transmission of the optical fiber was recovered as to be 6.7 \%-51.0 \% (given in Table \ref{tab1}) after the optical fiber was exposed to pulse laser light for 3 hours. The percentage of PB was less at 793 nm compared to the PB at short wavelengths of optical fiber. The reasons for less PB are;  a) Certain color center defects could not be bleached with an IR photon exposure (793 nm) due to different mechanisms of color center creation, which provided different qualities and color center energy levels b) The photon energy of the applied wavelength of 793 nm may be low \cite{Gebavi2012}. 

Consequently, the short wavelengths are more effective to the color centers, and the PB process in the short wavelengths is better than in the long wavelengths \cite{Henschel2000,Peng2010,Peng2017,Firstov2017}. Also, it was observed that the PB increased with the increasing energy of the pulsed laser light.  As seen in our results, the 532 nm pulsed laser provides better results because it transfers high energy compared to other wavelengths.

\vspace{-0.9em}

\subsection{Comparison of the literature and PB Mechanism}

The Yb-doped optical fibers have been used in high-power laser applications, and their performance depends on environmental effects such as radiation and high power. They are used for many hours in the applications. Therefore, the efficiency of the fiber decreases with time. We know that the cost of these systems is high, and they must be periodically changed due to the environmental effects. In the literature, some techniques have been investigated to recover damaged optical fibers, such as irradiation using the continuous laser \cite{Borman2016,Poulin2016}. The recovery levels are between 8 \%- 11 \% after 118 hours when the total dose is 1.45 kGy \cite{Borman2016, Poulin2016}. In the study of Henschel, and Köhn \cite{Henschel2000}, Ge-doped optical fiber showed 300 dB/km RIA after 10 kGy radiation and they used both an annealing process (28 hours) and a continuous laser with 830 nm wavelength. Finally, the RIA of the fiber decreased to 50 dB/km. Although the continuous laser is an effective process, the recovery can take a long time and increases with exposed radiation dose. In the present study, it has been observed that the laser system, which has a wide wave spectrum powered by the optical parametric amplifier (OPA) system, which is a new laser technique and can produce photons at a very high level in unit time compared to continuous laser light, shortens the recovery time and increases the recovery level at the same time. 
In our study, the pulsed laser system, which has a wide wave spectrum powered by the optical parametric amplifier (OPA) system, which is a new laser technique and can produce photons at a very high level in unit time compared to continuous laser light, shortens the recovery time and increases the recovery level at the same time (recovery levels of 30\%-96\% were achieved).

\section{Conclusion}
Yb-doped optical fibers are widely used in many applications like high-power fiber laser systems and they harsh radiation environments. We used the commercial Nufern optical fiber to analyze the radiation effect (10 kGy) and PB processes. The optical fibers were exposed to 10 kGy radiation and then recovered by the pulsed laser system, which has a wide wave spectrum powered by the optical parametric amplifier (OPA) system. The pulse laser shortens the recovery time and increases the recovery level at the same time and was used to recover the color centers in the study. Measurements were taken with a pulsed laser at 532 nm, 680 nm, and 793 nm, depending on the wavelength and exposure time. It was observed that the improvement increased as the exposure time of the Yb-doped optical fibers to the pulsed laser light increased. The recovery levels of color centers in the Yb doped optical fiber have been achieved to 30 \%-96 \%. In the optical fiber, the maximum PB was observed in the pulsed laser with a wavelength of 532 nm. In addition, it was observed that the PB in the optical fiber increased when the applied energy of the wavelength of the pulsed laser was increased.

\textbf{Acknowledgement} This work has been supported by the Scientific and Technological Research Council of Turkey, T\"{U}B\.{I}TAK, (No.120F281). We gratefully thank Dr. Atakan Tekgül for his constructive comments and Dr. Ali Karatutlu and Elif Yapar Yıldırım for the measurement of the doping concentration of Yb-doped optical fiber by using WDS.

\bibliography{sn-article}

\end{document}